\documentclass[spanish,english,aps,manuscript]{revtex4}
\usepackage[T1]{fontenc}
\usepackage[latin9]{inputenc}
\usepackage{amsthm}
\usepackage{amsmath}
\usepackage{graphicx}
\usepackage{amssymb}
\usepackage{esint}

\makeatletter
\@ifundefined{textcolor}{}
{%
 \definecolor{BLACK}{gray}{0}
 \definecolor{WHITE}{gray}{1}
 \definecolor{RED}{rgb}{1,0,0}
 \definecolor{GREEN}{rgb}{0,1,0}
 \definecolor{BLUE}{rgb}{0,0,1}
 \definecolor{CYAN}{cmyk}{1,0,0,0}
 \definecolor{MAGENTA}{cmyk}{0,1,0,0}
 \definecolor{YELLOW}{cmyk}{0,0,1,0}
 }

\makeatother

\usepackage{babel}
\addto\shorthandsspanish{\spanishdeactivate{~<>}}

\begin{document}

\title{Heat conduction in relativistic neutral gases revisited}

\author{A. L. García-Perciante , A. R. Méndez}

\address{Depto. de Matemáticas Aplicadas y Sistemas, Universidad Autónoma
Metropolitana-Cuajimalpa, Artificios 40 México D.F 01120, México}
\begin{abstract}
The kinetic theory of dilute gases to first order in the gradients
yields linear relations between forces and fluxes. The heat flux for
the relativistic gas has been shown to be related not only to the
temperature gradient but also to the density gradient in the representation
where number density, temperature and hydrodynamic velocity are the
independent state variables. In this work we show the calculation
of the corresponding transport coefficients from the full Boltzmann
equation and compare the magnitude of the relativistic correction. 
\end{abstract}
\maketitle

\section{Introduction}

It has been recently shown that, when constitutive equations to first
order in the gradients are introduced in the transport equations for
a relativistic gas, the system presents no intsability nor causality
issues \cite{jnnfm}. Moreover, the pathology identified by Hiscock
and Lindblom \cite{HL}, which in part lead to ruling out first order
theories, is due to the coupling of heat with acceleration. This coupling
has a phenomenological origin and is in contradiction with the results
obtained from relativistic kinetic theory. Due to this fact, the first
order theories are being currently reexamined and proposed as solid
frameworks from where one could extract the physics of high temperature
systems present both in astrophysical and experimental scenarios \cite{jnnfm}.

In this paper, the heat flux constitutive equation is obtained together
with the associated transport coefficients using the full collision
kernel in Boltzmann equation to first order in the gradients using
a representation where the density, hydrodynamic velocity and temperature
are the independent state variables. The heat flux in this scenario
is coupled with temperature and density gradients and two transport
coefficients are identified. The explicit form of such coeficients
are obtained for a constant scattering cross section model and the
results are shown to be consistent with the ones obtained by other
authors using a different representation \emph{only in the comoving
frame}. However, for an arbitrary observer, the stress energy tensor
includes Lorentz transformation factors \cite{ultimo}. The form of
such tensor, from which the transport equations are to be extracted,
is briefly discussed.

The present work is divided as follows. In Section II the theoretical
framework that sustains the calculation is presented. In Section III
the mathematical problem is set up as two separate integro-differential
equations whose solutions are formulated as expansions in orthogonal
polynomials from which the general form for the transport coefficients
is obtained. A constant scattering section model is assumed in Section
IV in order to calculate the collision integrals and compare the transport
coefficients as functions of $z$. The discussion of the results,
including their comparison with the solution obtained in Ref. \cite{ck-2}
as well as final remarks regarding the form of the stress-energy tensor
are included in Section V.

\section{Special relativistic Boltzmann equation}

Consider a neutral, dilute, single component, non-degenerate relativistic
fluid in the absence of external fields. Additionally, assume that
such a gas is characterized by values of the relativistic parameter
$z$, defined as the ratio of the thermal energy to the rest energy
a single particle, close to one. Much higher values of this parameter
correspond to very high temperature, ultra-relativistic gases for
which the neutrality of the particles here assumed may not be appropriate.
The non-relativistic gas corresponds to the limit where $z$ tends
to zero. Thus, for the system here considered, the space-time is given
by a Minkowski metric with a $+++-$ signature for which the position
and velocity four-vectors are given by \begin{equation}
x^{\nu}=\left(\vec{x},ct\right)\qquad v^{\nu}=\gamma_{\left(w\right)}\left(\vec{w},c\right)\label{eq:1}\end{equation}
 where $c$ is the speed of light and $\gamma_{\left(w\right)}=\left(1-w^{2}/c^{2}\right)^{-1/2}$.

The distribution function $f$ is an invariant such that $f\left(x^{\nu},v^{\nu}\right)d^{3}xd^{3}v$
, also an invariant, is the number of particles contained in a volume
in phase space. The relativistic Boltzmann equation for the evolution
of $f$ reads \cite{ck-2,degroot,Israel0-1}, \begin{equation}
v^{\alpha}f_{,\alpha}=J\left(f,f'\right)\label{eq:2}\end{equation}
 where the contraction on the left side corresponds to a total proper-time
derivative. That is, in the molecule's rest frame where $\vec{w}=\vec{0}$
\begin{equation}
v^{\alpha}f_{,\alpha}=\frac{v^{4}}{c}\frac{\partial f}{\partial\tau}=\frac{df}{d\tau}\label{eq:3}\end{equation}
 The right hand side of Eq. (\ref{eq:2}) is given by \begin{equation}
J\left(ff'\right)=\int\int\left\{ f\,'f_{1}\,'-f\, f_{1}\right\} \mathcal{F}\sigma\left(\Omega\right)d\Omega dv_{1}^{*}\label{eq:4}\end{equation}
 where $dv_{1}^{*}=d^{3}v_{1}/v_{1}^{4}$ and $\mathcal{F}$ is the
invariant flux given by \cite{ck-2} \begin{equation}
\mathcal{F}=\frac{1}{c^{2}}v^{4}v_{1}^{4}=\frac{1}{c}\sqrt{\left(v^{\alpha}v_{1\alpha}\right)^{2}-c^{4}}=\frac{1}{c}\sqrt{\left(\gamma_{(w)}\gamma_{(w_{1})}\left(\vec{w}\cdot\vec{w_{1}}-c^{2}\right)\right)^{2}-c^{4}}\label{eq:5}\end{equation}
 which reduces to the relative velocity in the non-relativistic limit.
The solution to the homogeneous Boltzmann equation is determined by
$J\left(ff'\right)=0$ together with the requirement of consistency
with the local equilibrium assumption. That is, thermodynamic equilibrium
is locally assumed and thus the state variables are given by \begin{equation}
N^{\nu}=\int f^{\left(0\right)}v^{\nu}dv^{*}\label{eq:6}\end{equation}
 \begin{equation}
\mathcal{T}^{\mu\nu}=\int f^{\left(0\right)}v^{\mu}v^{\nu}d^{*}v\label{eq:7}\end{equation}
 which are the particle flux and equilibrium stress-energy tensor
respectively. The thermodynamical, local equilibrium, variables for
the system can then be extracted from the previous tensors as \begin{equation}
n=-\frac{N^{\nu}\mathcal{U}_{\nu}}{c^{2}}\label{eq:8}\end{equation}
 \begin{equation}
n\varepsilon=\frac{\mathcal{U}_{\mu}\mathcal{U}_{\nu}}{c^{2}}\mathcal{T}^{\mu\nu}.\label{eq:9}\end{equation}
 It has been recently shown that the heat flux can be defined, as
in the non-relativistic case, as the average of the peculiar or chaotic
kinetic energy \cite{ultimo}. Thus, since this velocity is the one
measured by an observer locally comoving with the fluid element, the
calculation will be performed in a comoving frame. Therefore, the
fluid's hydrodynamic four velocity has only a temporal component:
\begin{equation}
\mathcal{U}^{\nu}=\left(\vec{0},\, c\right)\label{eq:10-1}\end{equation}
 The dissipative terms in the stress-energy tensor, as defined in
Ref. \cite{ultimo} from a standard tensor decomposition \cite{Eckart1-1},
include a four-vector that in an arbitrary frame can be calculated
as \begin{equation}
\tau^{\mu}=c^{2}L_{\nu}^{\mu}q^{\nu}\label{eq:11}\end{equation}
 where $q^{\nu}$ is the heat flux, calculated in the comoving frame,
and $L_{\nu}^{\mu}$ a Lorentz transfromation between the laboratory
and each local equilibrium element of the fluid. This idea was firstly
set forward, for the equilibrium quantities, by S. Weinberg \cite{weinberg}.
This result was obtained by introducing such transformation to relate
chaotic and molecular velocities and shows that the heat flux can
be consistently defined only in the comoving frame. Because of that,
from now on we will consider the hydrodynamic four-velocity as given
by Eq. (\ref{eq:10-1}) and the molecular velocity $v^{\mu}$ will
correspond to the chaotic velocity.

In order to solve Eq. (\ref{eq:2}) the standard Chapman-Enskog method
will be used \cite{ck-2,cc-1}. Thus, the solution is approximated
by \begin{equation}
f=f^{\left(0\right)}\left(1+\phi\left(v^{\mu}\right)\right)\label{eq:12}\end{equation}
 where $f^{(0)}$ is the Jüttner equilibrium distribution function
which, in the comoving frame, reads \cite{juttner} \begin{equation}
f^{\left(0\right)}=\frac{n}{4\pi c^{3}}\frac{1}{zK_{2}\left(\frac{1}{z}\right)}\exp^{-\frac{\gamma}{z}},\label{eq:13}\end{equation}
 Here $z=kT/mc^{2}$ is the relativistic parameter, where $T$ is
the local temperature, $k$ the Boltzmann constant, and $K_{n}$ is
the $n$-th modified Bessel function of the second kind. The solubility
conditions imposed on $\phi(\vec{v})$ are given by \begin{equation}
\int f^{\left(0\right)}\epsilon\phi\left(v^{\mu}\right)\left(\begin{array}{c}
m\gamma v^{4}\\
mv^{\nu}\end{array}\right)dv^{*}=0\label{eq:14}\end{equation}
 which amounts to restrict the local state variables to be defined
through the local equilibrium state. The proposed solution given in
Eq. (\ref{eq:12}) is substituted in Eq. (\ref{eq:2}). Considering
the deviation from the local equilibrium state $\phi(v^{\mu})$ to
be a first order quantity, one obtains a linearized first order Boltzmann
equation which can be written as \begin{equation}
v^{\alpha}f_{,\alpha}^{\left(0\right)}=f^{\left(0\right)}\mathbb{C}\left(\phi\left(v^{\mu}\right)\right)\label{lbe}\end{equation}
 where the linearized collision kernel is given by \begin{equation}
\mathbb{C}\left(\phi\right)=\int\int\left\{ \phi'_{1}+\phi'-\phi_{1}-\phi\right\} f_{1}^{\left(0\right)}\mathcal{F}\sigma\left(\Omega\right)d\Omega dv_{1}^{*}.\label{eq:15}\end{equation}
 The general solution to equation (\ref{lbe}) is given by the sum
of the homogeneous solution plus a particular solution, $\phi=\phi_{H}+\phi_{P}$.
The homogeneous solution is obtained as a linear combination of the
collision invariants \begin{equation}
\mathbb{C}\left(\begin{array}{c}
mv^{\mu}\\
m\gamma v^{4}\end{array}\right)=0.\label{eq:16}\end{equation}
 Existence of the particular solution is guaranteed by imposing an
orthogonality condition on the homogeneous solution and the inhomogeneous
equation namely, \begin{equation}
\int\left(\begin{array}{c}
mv^{\mu}\\
m\gamma v^{4}\end{array}\right)v^{\alpha}f_{,\alpha}^{\left(0\right)}dv^{*}=0,\label{eq:17}\end{equation}
 Equations (\ref{eq:17}) are the relativistic Euler equations obtained
through the equilibrium solution \cite{ch,pa08}. In the absence of
external forces, the left hand side of the relativistic Boltzmann
equation is written \begin{equation}
v^{\alpha}f_{,\alpha}=v^{\alpha}\left(\frac{\partial f^{\left(0\right)}}{\partial n}n_{,\alpha}+\frac{\partial f^{\left(0\right)}}{\partial T}T_{,\alpha}+\frac{\partial f^{\left(0\right)}}{\partial u^{\mu}}u_{;\alpha}^{\mu}\right).\label{left18}\end{equation}
 The next step consists in substituting the derivatives of the Jüttner
function and using the Euler equations to write the time derivatives
in terms of the gradients. Such equations constitute a closed set
for the state variables. At this point an appropriate representation
needs to be chosen and thus we consider $n,\, T$ and $\mathcal{U}^{\nu}$
as the set of state variables. The Euler equations are written as
\begin{equation}
\dot{n}=-nu_{;\alpha}^{\alpha}\label{eq:19}\end{equation}
 \begin{equation}
\dot{u}_{\alpha}=-zc^{2}\frac{K_{2}\left(\frac{1}{z}\right)}{K_{3}\left(\frac{1}{z}\right)}\left(\frac{n_{,\mu}}{n}+\frac{T_{,\mu}}{T}\right)h_{\alpha}^{\mu}\label{eq:20}\end{equation}
 \begin{equation}
\dot{T}=-\frac{T\beta}{nC_{n}k}u_{;\alpha}^{\alpha}\label{eq:21}\end{equation}
 where $\dot{\left(\right)}=u^{\nu}\left(\right)_{;\nu}$. Here the
gradient of the hydrostatic pressure has been written in terms of
the gradients of the number density and temperature by using an ideal
gas equation of state which can be easily shown to hold for dilute
special relativistic gases. We have also used the relation\begin{equation}
\frac{p}{zc^{2}\left(\frac{n\epsilon}{c^{2}}+\frac{p}{c^{2}}\right)}=\frac{K_{2}\left(\frac{1}{z}\right)}{K_{3}\left(\frac{1}{z}\right)}\label{eq:22}\end{equation}
 which can be verified by calculating $\varepsilon$ and $p$ from
the local equilibrium distribution function. After a somewhat tedieous
but straightforward algebraic manupulation one can write Eq. (\ref{lbe})
as follows \begin{equation}
v^{\beta}h_{\beta}^{\alpha}\left\{ \frac{n_{,\alpha}}{n}\left(1-\gamma\frac{K_{2}\left(\frac{1}{z}\right)}{K_{3}\left(\frac{1}{z}\right)}\right)+\frac{T_{,\alpha}}{T}\left(1+\frac{\gamma}{z}-\gamma\frac{K_{2}\left(\frac{1}{z}\right)}{K_{3}\left(\frac{1}{z}\right)}-\frac{K_{3}\left(\frac{1}{z}\right)}{zK_{2}\left(\frac{1}{z}\right)}\right)\right\} =\mathbb{C}\left(\phi\right)\label{inhomogenea1}\end{equation}
 where the term proportional to the hydrodynamic velocity gradient
does not arise since the calculations are performed in a comoving
frame.

The solution to Eq. (\ref{inhomogenea1}) is given by \begin{equation}
\phi=\mathcal{A}\left(\gamma\right)v^{\beta}h_{\beta}^{\alpha}\frac{T_{,\alpha}}{T}+\mathcal{B}\left(\gamma\right)v^{\beta}h_{\beta}^{\alpha}\frac{n_{,\alpha}}{n}+\alpha+\tilde{\alpha}_{\nu}v^{\nu}.\label{solution-2}\end{equation}
 The first two terms are the particular solution and the last two
terms correspond to the solution of the homogeneous equation. The
solubility conditions are thus written as\begin{equation}
\int\left(\mathcal{A}(\gamma)v^{\beta}h_{\beta}^{\alpha}\frac{T_{,\alpha}}{T}+\mathcal{B}(\gamma)v^{\beta}h_{\beta}^{\alpha}\frac{n_{,\alpha}}{n}+\alpha+\tilde{\alpha}_{\nu}v^{\nu}\right)\psi f^{(0)}dv^{*}=0\label{eq:23}\end{equation}
 where $\psi=mv^{\mu},m\gamma^{2}$. These conditions imply, as shown
in Appendix A, that the constant $\alpha$ vanishes and $\tilde{\alpha}_{\beta}$
is proportional to both $h_{\beta}^{\alpha}n_{,\alpha}$ and $h_{\beta}^{\alpha}T_{,\alpha}$
so that Eq. (\ref{solution-2}) reads \begin{equation}
\phi=\mathcal{A}(\gamma)v^{\beta}h_{\beta}^{\alpha}\frac{T_{,\alpha}}{T}+\mathcal{B}(\gamma)v^{\beta}h_{\beta}^{\alpha}\frac{n_{,\alpha}}{n}\label{eq:24}\end{equation}
 In Eq. (\ref{eq:24}), since $n_{,\alpha}$ and $T_{,\alpha}$ are
considered independent forces, $\mathcal{A}\left(\gamma\right)$ and
$\mathcal{B}\left(\gamma\right)$ are subject to the constraints \begin{equation}
\int\mathcal{A}(\gamma)\gamma^{2}\omega^{2}f^{(0)}dv^{*}=0,\label{eq:r1}\end{equation}
 \begin{equation}
\int\mathcal{B}(\gamma)\gamma^{2}\omega^{2}f^{(0)}dv^{*}=0.\label{eq:r2}\end{equation}
 To take full advantage of the fact that the unknowns $\mathcal{A}$
and $\mathcal{B}$ are functions of $\gamma$, we perform all integrals
in such variable using the relation \begin{equation}
dv^{*}=4\pi c^{3}\sqrt{\gamma^{2}-1}d\gamma\label{eq:25}\end{equation}
 which is obtained in Appendix B.

\section{Expansion in orthogonal polynomials}

By substituting Eq. (\ref{eq:24}) in Eq. (\ref{inhomogenea1}), the
mathematical problem set up in the previous section yields two independent
integral equations given by \begin{equation}
v^{\beta}h_{\beta}^{\alpha}\left\{ 1+\frac{\gamma}{z}-\gamma\frac{K_{2}\left(\frac{1}{z}\right)}{K_{3}\left(\frac{1}{z}\right)}-\frac{K_{3}\left(\frac{1}{z}\right)}{zK_{2}\left(\frac{1}{z}\right)}\right\} =\mathbb{C}(\mathcal{A}(\gamma)v^{\beta}h_{\beta}^{\alpha})\label{eq:26}\end{equation}
 and \begin{equation}
v^{\beta}h_{\beta}^{\alpha}\left\{ 1-\gamma\frac{K_{2}\left(\frac{1}{z}\right)}{K_{3}\left(\frac{1}{z}\right)}\right\} =\mathbb{C}(\mathcal{B}(\gamma)v^{\beta}h_{\beta}^{\alpha})\label{eq:27}\end{equation}
 subject to the contraints given by Eqs. (\ref{eq:r1}) and (\ref{eq:r2})
respectively. The unkown coefficients $\mathcal{A}$ and $\mathcal{B}$
are written in terms of orthogonal polynomials in $\gamma$ \begin{equation}
\mathcal{A}(\gamma)=\sum_{n=0}^{\infty}a_{n}\mathcal{L}_{n}(\gamma)\label{eq:28}\end{equation}
 \begin{equation}
\mathcal{B}(\gamma)=\sum_{n=0}^{\infty}b_{n}\mathcal{L}_{n}(\gamma)\label{eq:29}\end{equation}
 which satisfy the orthogonality condition \begin{equation}
\int\mathcal{L}_{n}(\gamma)\mathcal{L}_{m}(\gamma)p(\gamma)d\gamma=\delta_{nm},\label{eq:30}\end{equation}
 where the weight function $p(\gamma)=\exp^{-\frac{\gamma}{z}}(\gamma^{2}-1)^{3/2}$.
In the case where the hydrodynamic velocity is the one given in Eq.
(\ref{eq:10-1}), these polyomials, the first two of which are obtained
in Appendix C, are related to Kelly's set \cite{ck-2,kelly} $R_{\frac{3}{2}}^{n}$
by the relation \begin{equation}
R_{\frac{3}{2}}^{n}=\sqrt{3K_{2}\left(\frac{1}{z}\right)}z\mathcal{L}_{n}(\gamma)\label{eq:31}\end{equation}

In terms of the polynomials $\mathcal{L}$, the subsidiary conditions
can be written as \begin{equation}
\sum_{n=0}^{\infty}a_{n}\int\mathcal{L}_{n}(\gamma)p(\gamma)d\gamma=0,\label{eq:32}\end{equation}
 \begin{equation}
\sum_{n=0}^{\infty}b_{n}\int\mathcal{L}_{n}(\gamma)p(\gamma)d\gamma=0,\label{eq:33}\end{equation}
 Since $\mathcal{L}_{0}(\gamma)$ is constant, we have that $a_{0}=b_{0}=0$
and thus \begin{equation}
\mathcal{A}(\gamma)=\sum_{n=1}^{\infty}a_{n}\mathcal{L}_{n}(\gamma),\label{eq:34}\end{equation}
 \begin{equation}
\mathcal{B}(\gamma)=\sum_{n=1}^{\infty}b_{n}\mathcal{L}_{n}(\gamma).\label{eq:35}\end{equation}
 The heat flux in the Chapman-Enskog approximation, as clearly stated
in Ref. \cite{ultimo}, is given by the average of the chaotic kinetic
energy flux, a definition that encompasses the physical conception
of heat since the early developments of kinetic theory \cite{clausius,brush}.
Since in this work the molecular and chaotic velocities coincide,
we can write \begin{equation}
q^{\mu}=mc^{2}h_{\nu}^{\mu}\int\gamma v^{\nu}f^{(1)}d^{*}v\label{eq:36}\end{equation}
 or, subtituting Eq. (\ref{eq:36}) \begin{equation}
q^{\mu}=\frac{nm}{4\pi czK_{2}(\frac{1}{z})}\left[I_{(a)}^{\mu}+I_{(b)}^{\mu}\right].\label{eq:37}\end{equation}
 where \begin{equation}
I_{(a)}^{\mu}=h_{\nu}^{\mu}h_{\beta}^{\alpha}\frac{T_{,\alpha}}{T}\int\gamma v^{\nu}v^{\beta}\mathcal{A}(\gamma)e^{-\frac{\gamma}{z}}d^{*}v\label{eq:38}\end{equation}
 \begin{equation}
I_{(b)}^{\mu}=h_{\nu}^{\mu}h_{\beta}^{\alpha}\frac{n_{,\alpha}}{n}\int\gamma v^{\nu}v^{\beta}\mathcal{B}(\gamma)e^{-\frac{\gamma}{z}}d^{*}v.\label{eq:39}\end{equation}
 Notice that in both integrals only the $\nu,\beta=1,2,3$ terms survive
and from them, all $\nu\neq\beta$ ones also vanish because the integrands
are odd in the three-velocity. Thus, introducing Eqs. (\ref{eq:34})
and (\ref{eq:35}), the integrals read \begin{equation}
I_{(a)}^{\mu}=\frac{4\pi c^{5}}{3}h^{\mu\alpha}\frac{T_{,\alpha}}{T}\sum_{n=1}^{\infty}a_{n}\int\gamma\mathcal{L}_{n}(\gamma)p(\gamma)d\gamma\label{eq:40}\end{equation}
 \begin{equation}
I_{(b)}^{\mu}=\frac{4\pi c^{5}}{3}h^{\mu\alpha}\frac{n_{,\alpha}}{n}\sum_{n=1}^{\infty}b_{n}\int\gamma\mathcal{L}_{n}(\gamma)p(\gamma)d\gamma.\label{eq:41}\end{equation}
 As shown in Appendix C, we can write $\gamma=c_{0}\mathcal{L}_{0}(\gamma)+c_{1}\mathcal{L}_{1}(\gamma)$,
with $c_{1}=\sqrt{3g\left(z\right)}z$ where \begin{equation}
g\left(z\right)=5zK_{3}\left(\frac{1}{z}\right)+K_{2}\left(\frac{1}{z}\right)-\frac{K_{3}\left(\frac{1}{z}\right)^{2}}{K_{2}\left(\frac{1}{z}\right)}\label{eq:42}\end{equation}
 Using Eqs. (\ref{eq:40}-\ref{eq:42}) in the heat flux given by
Eq. (\ref{eq:37}) we obtain that \begin{equation}
q^{\mu}=-h^{\mu\alpha}\left[L_{T}\frac{T_{,\alpha}}{T}+L_{n}\frac{n_{,\alpha}}{n}\right]\label{eq:43}\end{equation}
 where the coefficients appearing in Eq. (\ref{eq:43}) are defined
as \begin{equation}
L_{T}=-\frac{nmc^{4}c_{1}}{3zK_{2}(\frac{1}{z})}a_{1}\label{eq:44}\end{equation}
 \begin{equation}
L_{n}=-\frac{nmc^{4}c_{1}}{3zK_{2}(\frac{1}{z})}b_{1}.\label{eq:45}\end{equation}

\section{Solution of the integral equations}

The coefficients $a_{1}$and $b_{1}$, in terms of which the coefficients
in Eqs. (\ref{eq:44}) and (\ref{eq:45}) are given, have to be obtained
from the solution of the integral equations (\ref{eq:26}) and (\ref{eq:27}).
In this section we outline such calculation to a first approximation.
The variational method used is the standard one as described in detail
in Ref. \cite{hirsh}.

First we notice that the integral equations (\ref{eq:26}-\ref{eq:27})
can be written as \begin{equation}
-\sqrt{3g\left(z\right)}z\left(\frac{K_{2}\left(\frac{1}{z}\right)}{K_{3}\left(\frac{1}{z}\right)}-\frac{1}{z}\right)f^{(0)}v^{\beta}h_{\beta}^{\alpha}\mathcal{L}_{1}(\gamma)=h_{\beta}^{\alpha}\sum_{n=1}^{\infty}a_{n}\mathbb{C}(\mathcal{L}_{n}(\gamma)v^{\beta}f^{(0)})\label{eq:46}\end{equation}
 \begin{equation}
-\frac{K_{2}\left(\frac{1}{z}\right)}{K_{3}\left(\frac{1}{z}\right)}\sqrt{3g\left(z\right)}zf^{(0)}v^{\beta}h_{\beta}^{\alpha}\mathcal{L}_{1}(\gamma)=h_{\beta}^{\alpha}\sum_{n=1}^{\infty}b_{n}\mathbb{C}(\mathcal{L}_{n}(\gamma)v^{\beta}f^{(0)})\label{eq:47}\end{equation}
 such that both have a similar structure. Indeed, since the dependence
on $\gamma$ on both is the same, the procedure only needs to be carried
out once for one of the equations and the solution for the other one
can be readily inferred by adjusting the dependence on the parameter
$z$. This similarity is consistent with the calculation in Ref. \cite{ck-2}
where only one integral equation needs to be solved for the coefficient
of a generalized thermodynamic force which includes contributions
from $\nabla T$ and $\nabla p$ in a single term.

Following the presciption mentioned above, we will only deal with
Eq. (\ref{eq:47}). Multiplying it by $\mathcal{L}_{m}(\gamma)v_{\nu}h_{\alpha}^{\nu}$
and integrating on both sides \begin{equation}
-\frac{\sqrt{3g}}{K_{3}\left(\frac{1}{z}\right)}\frac{n}{4\pi c^{3}}\int h_{\beta}^{\nu}e^{-\frac{\gamma}{z}}v_{\nu}v^{\beta}\mathcal{L}_{1}\mathcal{L}_{m}d^{*}v=h_{\beta}^{\nu}\sum_{n=1}^{\infty}b_{n}\int\mathcal{L}_{m}v_{\nu}\mathbb{C}(\mathcal{L}_{n}v^{\beta}f^{(0)})d^{*}v\label{eq:48}\end{equation}
 where we have omitted the $z$ and $\gamma$ dependences to short
notation. For the integral on left hand side, using that $e^{-\frac{\gamma}{z}}h_{\beta}^{\nu}v_{\nu}v^{\beta}=4\pi c^{5}p(\gamma)d\gamma$,
we have \begin{equation}
\int e^{-\frac{\gamma}{z}}h_{\beta}^{\nu}v_{\nu}v^{\beta}\mathcal{L}_{1}(\gamma)\mathcal{L}_{m}(\gamma)d^{*}v=4\pi c^{5}\delta_{1m}\label{eq:49}\end{equation}
 and thus, defining the collision brakett in the standard way \begin{equation}
\left[G,H\right]=-\frac{1}{n^{2}}\int G_{\alpha}\cdot\left[H_{\alpha1}^{'}+H_{\alpha}^{'}-H_{\alpha1}-H_{\alpha}\right]f^{\left(0\right)}f_{1}^{\left(0\right)}\mathcal{F}\sigma\left(\Omega\right)d\Omega dv_{1}^{*}dv^{*}\label{eq:50}\end{equation}
 in this equation $G_{\alpha}=G_{\alpha}(v_{\beta})$ and we have
used $H_{\alpha1}^{'}$ to denote $H_{\alpha}\left(v_{\beta1}^{'}\right)$.
The integral equations can be written as \begin{equation}
h_{\beta}^{\nu}\sum_{n=1}^{\infty}a_{n}\left[\mathcal{L}_{m}(\gamma)v_{\nu},\mathcal{L}_{n}(\gamma)v^{\beta}\right]=\left(\frac{K_{2}\left(\frac{1}{z}\right)}{K_{3}\left(\frac{1}{z}\right)}-\frac{1}{z}\right)\frac{c^{2}\sqrt{3g\left(z\right)}}{nK_{2}\left(\frac{1}{z}\right)}\delta_{1m}\label{eq:51}\end{equation}
 \begin{equation}
h_{\beta}^{\nu}\sum_{n=1}^{\infty}b_{n}\left[\mathcal{L}_{m}(\gamma)v_{\nu},\mathcal{L}_{n}(\gamma)v^{\beta}\right]=c^{2}\frac{\sqrt{3g\left(z\right)}}{nK_{3}\left(\frac{1}{z}\right)}\delta_{1m}\label{eq:52}\end{equation}
 By following the standard variational method \cite{hirsh}, the first
approximation for $a_{1}$ and $b_{1}$ can be shown to be given by
\begin{equation}
a_{1}=\left(\frac{K_{2}\left(\frac{1}{z}\right)}{K_{3}\left(\frac{1}{z}\right)}-\frac{1}{z}\right)\frac{\sqrt{3g\left(z\right)}}{nK_{2}\left(\frac{1}{z}\right)}c^{2}\left\{ h_{\beta}^{\nu}\left[\mathcal{L}_{1}(\gamma)v_{\nu},\mathcal{L}_{1}(\gamma)v^{\beta}\right]\right\} ^{-1}\label{eq:53}\end{equation}
 \begin{equation}
b_{1}=c^{2}\frac{\sqrt{3g\left(z\right)}}{nK_{3}\left(\frac{1}{z}\right)}\left\{ h_{\beta}^{\nu}\left[\mathcal{L}_{1}(\gamma)v_{\nu},\mathcal{L}_{1}(\gamma)v^{\beta}\right]\right\} ^{-1}\label{eq:54}\end{equation}
 Thus, in order to calculate the coefficients $a_{1}$ and $b_{1}$
to this level of approximation only one collision integral needs to
be calculated namely, $\left[\mathcal{L}_{1}(\gamma)v_{\nu},\mathcal{L}_{1}(\gamma)v^{\beta}\right]$.
In order to calculate such brakett, the well known identity \begin{align}
\left[G_{\alpha},H_{\beta}\right] & =\frac{1}{4n^{2}}\int\left[G_{\alpha}\left(v^{\mu'}\right)+G_{\alpha}\left(v_{1}^{\mu'}\right)-G_{\alpha}\left(v^{\mu}\right)-G_{\alpha}\left(v_{1}^{\mu}\right)\right]\cdot\label{eq:55}\\
 & \left[H_{\beta}\left(v_{1}^{\mu'}\right)+H_{\beta}\left(v^{\mu'}\right)-H_{\beta}\left(v_{1}^{\mu}\right)-H_{\beta}\left(v^{\mu}\right)\right]f^{\left(0\right)}f_{1}^{\left(0\right)}\mathcal{F}\sigma\left(\Omega\right)d\Omega dv_{1}^{*}dv^{*},\nonumber \end{align}
 will be used. Using also the momentum conservation law for collisions
and after several algebraic steps one can show that \begin{equation}
h_{\beta}^{\nu}\left[\mathcal{L}_{1}(\gamma)v_{\nu},\mathcal{L}_{1}(\gamma)v^{\beta}\right]=-\frac{\mathcal{I}_{1}-\mathcal{I}_{2}}{3g\left(z\right)z^{2}n^{2}m^{4}c^{5}}\label{eq:56}\end{equation}
 where the integrals $\mathcal{I}_{1}$ and $\mathcal{I}_{2}$ are
defined as\begin{equation}
\mathcal{I}_{1}=m^{4}c^{7}\int\int\gamma^{2}\left[\left(\gamma^{2}\right)'_{1}+\left(\gamma^{2}\right)'-\left(\gamma^{2}\right)_{1}-\gamma^{2}\right]f^{\left(0\right)}f_{1}^{\left(0\right)}\mathcal{F}\sigma d\Omega dv_{1}^{*}dv^{*}\label{eq:57}\end{equation}
 \begin{equation}
\mathcal{I}_{2}=-m^{4}c^{5}\int\int\gamma v^{\beta}\left[\left(\gamma v_{\beta}\right)'_{1}+\left(\gamma v_{\beta}\right)'-\left(\gamma v_{\beta}\right)_{1}-\gamma v_{\beta}\right]f^{\left(0\right)}f_{1}^{\left(0\right)}\mathcal{F}\sigma d\Omega dv_{1}^{*}dv^{*}.\label{eq:58}\end{equation}
 Thus, the general expressions for the coefficients $L_{T}$ and $L_{n}$
are given by \begin{equation}
L_{T}=\frac{3k^{2}T^{2}n^{2}m^{3}c^{7}}{\mathcal{I}_{1}-\mathcal{I}_{2}}\left(\frac{K_{2}\left(\frac{1}{z}\right)}{K_{3}\left(\frac{1}{z}\right)}-\frac{1}{z}\right)\left(\frac{g\left(z\right)}{K_{2}\left(\frac{1}{z}\right)}\right)^{2}\label{44a}\end{equation}
 \begin{equation}
L_{n}=-\frac{3k^{2}T^{2}n^{2}m^{3}c^{7}}{\mathcal{I}_{1}-\mathcal{I}_{2}}\frac{\left(g\left(z\right)\right)^{2}}{K_{3}\left(\frac{1}{z}\right)K_{2}(\frac{1}{z})}.\label{45a}\end{equation}
 In order to evaluate the integrals in $\mathcal{I}_{1}$ and $\mathcal{I}_{2}$,
a particular collision model needs to be proposed. In the next section
the simplest collision model, namely a constant cross section, will
be assumed in order to obtain expression for the coefficients and
asess their relative magnitude.

\section{Constant scattering cross section}

\begin{figure}[h]
\begin{centering}
\includegraphics[width=8.7cm]{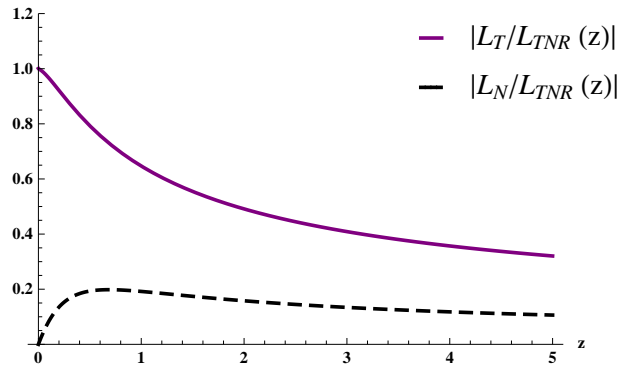} 
\par\end{centering}

\caption{The magnitudes of both coefficients, normalized to $L_{TNR}$ are
shown as functions of $z$.}

\label{fig1} 
\end{figure}

The simplest model that one might consider in order to calculate collision
integrals consists in assuming a constant cross section. The details
of the calculations for such model can be found in Ref. \cite{ck-2}.
Here, we will only quote the final results for the integrals in Eqs.
(\ref{eq:57}) and (\ref{eq:58}) \begin{equation}
\mathcal{I}_{1}=-\frac{64\pi n^{2}k^{6}T^{6}\sigma}{m^{2}c^{4}\left(K_{2}\left(\frac{1}{z}\right)\right)^{2}}\left(2K_{2}\left(\frac{2}{z}\right)+\frac{1}{z}K_{3}\left(\frac{2}{z}\right)\right)\label{eq:59}\end{equation}
 \begin{equation}
\mathcal{I}_{2}=\frac{64\pi n^{2}k^{6}T^{6}\sigma}{m^{2}c^{4}\left(K_{2}\left(\frac{1}{z}\right)\right)^{2}}\left(\frac{4}{z}K_{3}\left(\frac{2}{z}\right)+\frac{1}{z^{2}}K_{2}\left(\frac{2}{z}\right)\right)\label{eq:60}\end{equation}
 where $\sigma$ is the constant scattering cross section. Subtituting
these expressions in Eqs. (\ref{44a}) and (\ref{45a}) one obtains
\begin{equation}
L_{T}=-\frac{3ckT}{64\pi\sigma}\left(z\frac{K_{2}\left(\frac{1}{z}\right)}{K_{3}\left(\frac{1}{z}\right)}-1\right)\frac{\left(K_{2}\left(\frac{1}{z}\right)\left[\frac{1}{z}+5\mathcal{G}\left(\frac{1}{z}\right)-\frac{1}{z}\mathcal{G}\left(\frac{1}{z}\right)^{2}\right]\right)^{2}}{z^{4}\left(\frac{5}{z}K_{3}\left(\frac{2}{z}\right)+\left(\frac{1}{z^{2}}+2\right)K_{2}\left(\frac{2}{z}\right)\right)}\label{eq:61}\end{equation}
 \begin{equation}
L_{n}=-\frac{3ckT}{64\pi\sigma}\frac{K_{2}\left(\frac{1}{z}\right)\left(K_{2}\left(\frac{1}{z}\right)\left[\frac{1}{z}+5\mathcal{G}\left(\frac{1}{z}\right)-\frac{1}{z}\mathcal{G}\left(\frac{1}{z}\right)^{2}\right]\right)^{2}}{z^{3}K_{3}\left(\frac{1}{z}\right)\left(\frac{5}{z}K_{3}\left(\frac{2}{z}\right)+\left(\frac{1}{z^{2}}+2\right)K_{2}\left(\frac{2}{z}\right)\right)}\label{eq:62}\end{equation}
 which are to some extent approximations to the coefficients in appearing
in the heat flux constitutve equation in the $n,\, T,\,\mathcal{U}^{\nu}$
representation. In the limit of small relativistic paramenter $z$
one recovers the non-relativistic values \begin{equation}
L_{T}\sim L_{TNR}\left\{ 1-\frac{3}{16}z+....\right\} \label{eq:63}\end{equation}
 \begin{equation}
L_{n}\sim-L_{TNR}\left\{ z-\frac{27}{16}z^{2}+...\right\} \label{eq:64}\end{equation}
 where $L_{TNR}=(75mc^{3}/256\sqrt{\pi}\sigma)z^{3/2}$ is the usual
non-relativistic thermal conductivity for hard spheres divided by
$T$. Notice that indeed, as $z$ goes to zero $L_{T}\rightarrow L_{TNR}$
and $L_{n}\rightarrow0$. In Fig. \ref{fig1} the magnitudes of both
coefficients, normalized to $L_{TNR}$ are shown as functions of $z$.

\section{Discussion and final remarks}

In the previous section the transport coefficients involved in the
constitutive equation for the heat flux in terms of the forces $\nabla T$
and $\nabla n$ have been obtained. The calculation has been performed
in the fluid's local comoving frame, that is in terms of the chaotic
or pecular velocity. The Chapman-Enskog solution method of Boltzmann's
relativistic equation in this representation leads to two independent
integral equations. Once the solution is written in terms of a suitable
orthogonal set of polynomials, both equations have the same structure
and the solution is finally obtained using the standard variational
method. By assuming a very simple constant cross section model to
calculate the collision integrals, the coefficients can be studied.
The results are shown in Fig. 1. It can be clearly seen that for small
values of $z$ the non-relativistic limit is verified. For larger
values of the relativistic parameter the coefficient $L_{n}$ increases
and has the same order of magnitude of $L_{T}$. That is, if the gradients
of the system are of the same order of magnitude the relativistic
effect $q\propto\nabla n$ is as important as the usual Fourier effect.

In previous work, Cercigniani \& Kremer obtained the constitutive
equation for the heat flux in a relativistic neutral gas in a different
way. The definition they used for the heat flux is the one obtained
from the phenomenology using Eckart's decomposition in an arbitary
frame. However, in the comoving frame both definitions, the one given
here in Eq. (\ref{eq:36}) and the one they consider namely,

\begin{equation}
q^{\alpha}=h_{\beta}^{\alpha}\mathcal{U}_{\gamma}c\int p^{\beta}p^{\gamma}f\frac{d^{3}p}{p^{4}}\label{eq:65}\end{equation}
 where $p^{\alpha}$ is the four-momentum, are the same. However,
after the calculation of the heat flux from the Chapman-Enskog solution
of the Boltzman equation, in Ref. \cite{ck-2} the heat flux is obtained
in terms of a relativistic thermal force which includes both the Fourier
term as well as the relativistic correction which they leave in terms
of the gradient of the hydrostatic pressure. Their result is\begin{equation}
q^{\alpha}=\lambda h^{\alpha\beta}\left[T_{,\beta}-\frac{T}{nh_{e}}p_{,\beta}\right]\label{eq:66}\end{equation}
 where\begin{equation}
h_{E}=\epsilon+\frac{p}{n}=\frac{p}{nz}\frac{K_{3}\left(\frac{1}{z}\right)}{K_{2}\left(\frac{1}{z}\right)}\label{eq:67}\end{equation}
 and $\lambda$ is given by\begin{equation}
\lambda=-\frac{3kp^{2}m^{2}c^{5}\left(\frac{1}{z}+5\mathcal{G}\left(\frac{1}{z}\right)-\frac{1}{z}\mathcal{G}\left(\frac{1}{z}\right)^{2}\right)^{2}}{I_{1}-c^{2}I_{2}}\label{eq:68}\end{equation}
 with the integrals $I_{1}$ and $I_{2}$ defined as \begin{align*}
I_{1}= & \mathcal{U}_{\alpha}\mathcal{U}_{\beta}\mathcal{U}_{\gamma}\mathcal{U}_{\delta}\times\\
 & \iint p^{\alpha}p^{\beta}\left\{ \left(p^{\gamma}p^{\delta}\right)'_{1}+\left(p^{\gamma}p^{\delta}\right)'-\left(p^{\gamma}p^{\delta}\right)_{1}-p^{\gamma}p^{\delta}\right\} f_{c}^{\left(0\right)}f_{c1}^{\left(0\right)}F\sigma d\Omega\frac{d^{3}p_{1}}{p_{1}^{4}}\frac{d^{3}p}{p^{4}}\end{align*}
 \begin{align*}
I_{2}= & \mathcal{U}_{\alpha}\mathcal{U}_{\gamma}\times\\
 & \iint p^{\alpha}p^{\beta}\left\{ \left(p^{\gamma}p_{\beta}\right)'_{1}+\left(p^{\gamma}p_{\beta}\right)'-\left(p^{\gamma}p_{\beta}\right)_{1}-p^{\gamma}p_{\beta}\right\} f_{c}^{\left(0\right)}f_{c1}^{\left(0\right)}F\sigma d\Omega\frac{d^{3}p_{1}}{p_{1}^{4}}\frac{d^{3}p}{p^{4}}\end{align*}
 The invariant flux in the expressions above is related with the one
in Eq. (\ref{eq:4}) by $F=m^{2}c\mathcal{F}$ and also the distribution
function in Ref. \cite{ck-2} differs by a factor from the one used
in this work: $f_{c}=\frac{1}{m^{3}}f$.

In order to compare Eq. (\ref{eq:66}) with our results the hydrostatic
pressure can be written in terms of $\nabla n$ and $\nabla T$ by
means of the ideal gas equation of state. Also the hydrodynamic velocity
has to be substituted by the one given in Eq. (\ref{eq:10-1}), which
accounts to writting Cercigniani's result \cite{ck-2} in the comoving
frame. After these changes have been introduced, the integration variable
changed to $\gamma$ and the different signature is taken into account,
one can readily verify that $I_{1}=\mathcal{I}_{1}$ and $I_{2}=c^{-2}\mathcal{I}_{2}$
and that the constitutive equation in Eq. (\ref{eq:66}) can be written
as\begin{equation}
q^{\alpha}=h^{\alpha\beta}\left[\lambda T\left(1-z\frac{K_{2}\left(\frac{1}{z}\right)}{K_{3}\left(\frac{1}{z}\right)}\right)\frac{T_{,\beta}}{T}-\lambda Tz\frac{K_{2}\left(\frac{1}{z}\right)}{K_{3}\left(\frac{1}{z}\right)}\left(\frac{n_{,\beta}}{n}\right)\right]\label{71}\end{equation}
 which has the same structure of Eq. (\ref{eq:43}). Subtituting the
value for $\lambda$ obtained in Ref. \cite{ck-2} which can be written
as\begin{equation}
\lambda=-\frac{3kn^{2}m^{4}c^{9}}{I_{1}-c^{2}I_{2}}\left(\frac{g\left(z\right)}{K_{2}\left(\frac{1}{z}\right)}\right)^{2}\label{eq:72}\end{equation}
 yields\begin{equation}
\lambda T\left(1-z\frac{K_{2}\left(\frac{1}{z}\right)}{K_{3}\left(\frac{1}{z}\right)}\right)=\left(\frac{K_{2}\left(\frac{1}{z}\right)}{K_{3}\left(\frac{1}{z}\right)}-\frac{1}{z}\right)\frac{3n^{2}k^{2}T^{2}m^{3}c^{7}}{I_{1}-c^{2}I_{2}}\left(\frac{g\left(z\right)}{K_{2}\left(\frac{1}{z}\right)}\right)^{2}\label{eq:73}\end{equation}
 \begin{equation}
\lambda Tz\frac{K_{2}\left(\frac{1}{z}\right)}{K_{3}\left(\frac{1}{z}\right)}=-\frac{3n^{2}k^{2}T^{2}m^{3}c^{7}}{I_{1}-c^{2}I_{2}}\frac{\left(g\left(z\right)\right)^{2}}{K_{3}\left(\frac{1}{z}\right)K_{2}\left(\frac{1}{z}\right)}\label{eq:74}\end{equation}
 which are precisely the coefficients in Eqs. (\ref{44a}) and (\ref{45a}).
Thus, one concludes that both calculations are consistent in the comoving
frame.

As mentioned above, the constitutive equation obtained in this work
is equivalent to the one obtained in Ref. \cite{ck-2} only in the
comoving frame since in that case the definitions for the heat flux
are identical. In this work we follow the ideas set forth in Ref.
\cite{ultimo} where the heat flux is a local quantity which only
makes sense in the comoving frame where the molecular velocity coincides
with the peculiar velocity. However, the term appearing in the stress-energy
tensor and that will ultimately impact the transport equations is
in our case given by $\tau^{\mu}=c^{2}L_{\nu}^{\mu}q^{\nu}$ with
$q^{\nu}$ given by Eq. (\ref{eq:36}) while in the traditional calculations
what is obtained is $\tau^{\mu}=q^{\nu}$ with $q^{\nu}$ given by
Eq. (\ref{eq:65}). This difference is discussed in Ref. \cite{ultimo}
and its implications will be explored in future work.
\begin{acknowledgments}
The authors wish to thank Dr. A. Sandoval-Villalbazo, Prof. García-Colín
and Prof. G. Medeiros-Kremer for valuable comments and suggestions.
This work was partially supported by PROMEP grant UAM-PTC-142.
\end{acknowledgments}

\section*{Appendix A}

In this appendix we show that Eq. (\ref{eq:24}) is valid in view
of the subsidiary conditions given in Eq. (\ref{eq:23}). Equation
(\ref{eq:23}) can be written as \begin{equation}
\int\left(\mathcal{A}(\gamma)v^{\ell}\frac{T_{,\ell}}{T}+\mathcal{B}(\gamma)v^{\ell}\frac{n_{,\ell}}{n}+\alpha+\tilde{\alpha}_{\ell}v^{\ell}+\tilde{\alpha}_{4}v^{4}\right)\psi f^{(0)}dv^{*}=0,\label{eq:75}\end{equation}
 where we have separated spatial and temporal terms in the contraction
$\tilde{\alpha}_{\nu}v^{\nu}$ since they have opposite parity.

For $\psi=v^{4}=\gamma c$ the first, second and forth terms yield
odd integrands and thus \begin{equation}
\int\left(\alpha+\tilde{\alpha}_{4}v^{4}\right)\gamma f^{(0)}dv^{*}=0\label{eq:76}\end{equation}
 For $\psi=v^{\ell}$ with $\ell=1,\,2,\,3$ the third and fifth terms
do not contribute because of their odd parity and we obtain \begin{equation}
\int\left(\mathcal{A}(\gamma)\frac{T,_{\ell}}{T}+\mathcal{B}(\gamma)\frac{n,_{\ell}}{n}+\tilde{\alpha}_{\ell}\right)v^{\ell}v^{k}f^{(0)}dv^{*}=0,\label{77}\end{equation}
 which vanishes for $\ell\neq k$. Thus \begin{equation}
\int\left(\mathcal{A}\frac{T,_{\ell}}{T}+\mathcal{B}\frac{n,_{\ell}}{n}+\tilde{\alpha}_{\ell}\right)v^{\ell}v^{k}f^{(0)}dv^{*}=\frac{\delta^{k\ell}}{3}\int\left(\mathcal{A}\frac{T,_{\ell}}{T}+\mathcal{B}\frac{n,_{\ell}}{n}+\tilde{\alpha}_{\ell}\right)\gamma^{2}w^{2}f^{(0)}dv^{*},\label{eq:78}\end{equation}
 and thus the condition in Eq. (\ref{77}) reduces to \begin{equation}
\int\left(\mathcal{A}(\gamma)\frac{T,_{\ell}}{T}+\mathcal{B}(\gamma)\frac{n,_{\ell}}{n}+\tilde{\alpha}_{\ell}\right)\gamma^{2}w^{2}f^{(0)}dv^{*}=0.\label{eq:79}\end{equation}
 Finally, for $\psi=\gamma^{2}$ the parity in the terms is the same
as for $\psi=v^{4}$ such that \begin{equation}
\int\left(\alpha+\tilde{\alpha}_{4}v^{4}\right)\gamma^{2}f^{(0)}dv^{*}=0\label{eq:80}\end{equation}
 From equation (\ref{77}) we have that \begin{equation}
\tilde{\alpha}_{k}=-\frac{T,_{k}}{T}\frac{\int\mathcal{A}(\gamma)\gamma^{2}\omega^{2}f^{(0)}dv^{*}}{\int\gamma^{2}\omega^{2}f^{(0)}dv^{*}}-\frac{n,_{k}}{n}\frac{\int\mathcal{B}(\gamma)\gamma^{2}\omega^{2}f^{(0)}dv^{*}}{\int\gamma^{2}\omega^{2}f^{(0)}dv^{*}},\label{81}\end{equation}
 such that we can redefine \begin{equation}
\mathcal{A}(\gamma)v^{k}\frac{T,_{k}}{T}+\mathcal{B}(\gamma)v^{k}\frac{n,_{k}}{n}+\tilde{\alpha}_{k}v^{k}\longrightarrow a(\gamma)v^{k}\frac{T,_{k}}{T}+b(\gamma)v^{k}\frac{n,_{k}}{n}\label{eq:82}\end{equation}
 and now the subsidiary condition (\ref{77}) reads \begin{align}
\int\left[\mathcal{A}(\gamma)\frac{T,_{k}}{T}+\mathcal{B}(\gamma)\frac{n,_{k}}{n}\right]\gamma^{2}\omega^{2}f^{(0)}dv^{*} & =0.\label{int-indep}\end{align}
 Equations (\ref{eq:76}) and (\ref{eq:80}) can be written as an
homogeneous system for $(\alpha,\tilde{\alpha_{4}})$: \begin{align}
\alpha g_{11}+\tilde{\alpha_{4}}g_{12} & =0,\nonumber \\
\alpha g_{21}+\tilde{\alpha_{4}}g_{22} & =0,\label{eq:83}\end{align}
 where \begin{align}
g_{11}=\int\gamma f^{(0)}dv^{*}, & \quad g_{12}=g_{21}=\int\gamma^{2}cf^{(0)}dv^{*},\quad & g_{22}=\int\gamma^{3}c^{2}f^{(0)}dv^{*},\label{eq:84}\end{align}
 such that the determinant does not vanish and the solution is the
trival one $\alpha=\tilde{\alpha_{4}}=0$. Putting toghether these
two results, we can conclude that the proposed solution consistent
with the subsiadiary conditions is the one given by Eq. (\ref{eq:24})
where the condition in Eq. (\ref{77}) still needs to be enforced.
Considering both forces $T_{,k}$ and $n_{,k}$ as independent forces,
this requirement is written as two separate conditions in Eqs. (\ref{eq:r1})
and (\ref{eq:r2}).

\section*{Appendix B}

In this appendix, the identity $dv^{*}=4\pi c^{3}\sqrt{\gamma^{2}-1}d\gamma$,
which is used in several parts of the work, will be obtained. The
invariant volume element in velocity space is given by \cite{degroot,ck-2}
\begin{equation}
dv^{*}=c\frac{dv^{3}}{v^{4}}=\frac{dv^{3}}{\gamma}\label{eq:85}\end{equation}
 or, in terms of the three-velocity $\vec{w}$ \begin{equation}
dv^{*}=\det\left[J\right]\frac{d^{3}w}{\gamma}\label{eq:86}\end{equation}
 where the Jacobian matrix has components $J_{ij}=\partial v_{i}/\partial w_{j}$
which can be shown to be given by \begin{equation}
J_{ij}=\gamma\left(\delta_{ij}+\gamma^{2}\frac{w_{i}w_{j}}{c^{2}}\right)\label{eq:87}\end{equation}
 Using the identity $\det\left[\delta_{ij}+A_{i}B_{j}\right]=1+A_{i}B^{i}$
\begin{equation}
\det\left[J\right]=\gamma^{3}\left(1+\gamma^{2}\frac{w^{2}}{c^{2}}\right)=\gamma^{5}\label{eq:88}\end{equation}
 and thus, introducing spherical coordinates for $\vec{w}$ and that
$w^{2}dw=\frac{c^{3}}{\gamma^{4}}\sqrt{\gamma^{2}-1}d\gamma$ we finally
obtain \begin{equation}
dv^{*}=4\pi c^{3}\sqrt{\gamma^{2}-1}d\gamma\label{eq:89}\end{equation}

\section*{Appendix C}

The orthogonal polynomials are obtained using the standard Gram-Schmidt
procedure \cite{arfken}. The proposed polynomials are $\mathcal{L}_{n}(\gamma)=a_{0n}+a_{1n}\gamma+a_{2n}\gamma^{2}+\dots+a_{nn}\gamma^{n}$
which are required to satisfy the orthonormality condition \begin{equation}
\int\mathcal{L}_{n}(\gamma)\mathcal{L}_{m}(\gamma)p(\gamma)d\gamma=\delta_{mn}\label{orto}\end{equation}
 where $p\left(\gamma\right)=\exp^{-\frac{\gamma}{z}}(\gamma^{2}-1)^{3/2}$.
For $n=0$ we have $\mathcal{L}_{0}(\gamma)=a_{00}$ and the orthonormality
condition (\ref{orto}) yields \begin{equation}
\mathcal{L}_{0}(\gamma)=\frac{1}{\sqrt{3}zK_{2}(\frac{1}{z})^{\frac{1}{2}}}.\label{eq:90}\end{equation}
 For $n=1$ we have $\mathcal{L}_{1}(\gamma)=a_{01}+a_{11}\gamma$
and equation (\ref{orto}) yields

\begin{equation}
a_{01}=-\frac{K_{3}\left(\frac{1}{z}\right)}{K_{2}\left(\frac{1}{z}\right)}\frac{1}{\sqrt{3}z}\left[5zK_{3}\left(\frac{1}{z}\right)+K_{2}\left(\frac{1}{z}\right)-\frac{K_{3}\left(\frac{1}{z}\right)^{2}}{K_{2}\left(\frac{1}{z}\right)}\right]^{-\frac{1}{2}}\label{eq:91}\end{equation}
 \begin{equation}
a_{11}=\frac{1}{\sqrt{3}z}\left[5zK_{3}\left(\frac{1}{z}\right)+K_{2}\left(\frac{1}{z}\right)-\frac{K_{3}\left(\frac{1}{z}\right)^{2}}{K_{2}\left(\frac{1}{z}\right)}\right]^{-\frac{1}{2}}\label{eq:92}\end{equation}
 if we define $g\left(z\right)=5zK_{3}\left(\frac{1}{z}\right)+K_{2}\left(\frac{1}{z}\right)-\frac{K_{3}\left(\frac{1}{z}\right)^{2}}{K_{2}\left(\frac{1}{z}\right)}$
\begin{equation}
\mathcal{L}_{1}(\gamma)=\frac{1}{\sqrt{3g\left(z\right)}z}\left[-\frac{K_{3}\left(\frac{1}{z}\right)}{K_{2}\left(\frac{1}{z}\right)}+\gamma\right].\label{93}\end{equation}
 We only need these two polynomials in order to calculate the heat
flux. Additionally, the coefficients in $\gamma=c_{0}\mathcal{L}_{0}+c_{1}\mathcal{L}_{1}$
are introduced in Sect. III. A simple way of obtaining them is to
solve Eq. (\ref{93}) for $\gamma$ and use Eq. (\ref{eq:90}) which
yields \begin{equation}
\gamma=\sqrt{3g\left(z\right)}z\mathcal{L}_{1}(\gamma)+\sqrt{3}z\frac{K_{3}\left(\frac{1}{z}\right)}{\sqrt{K_{2}\left(\frac{1}{z}\right)}}\mathcal{L}_{0}\label{eq:94}\end{equation}
 from which $c_{0}=\sqrt{3}z\frac{K_{3}\left(\frac{1}{z}\right)}{\sqrt{K_{2}\left(\frac{1}{z}\right)}}$
and $c_{1}=\sqrt{3g\left(z\right)}z$.

\end{document}